\documentclass[conference,a4paper]{IEEEtran}

\usepackage[utf8]{inputenc} 
\usepackage[T1]{fontenc}
\usepackage{url}
\usepackage{ifthen}
\usepackage{cite}
\usepackage[cmex10]{amsmath} 

\newtheorem{theorem}{Theorem}

\newtheorem{lem}{Lemma}

\ifCLASSINFOpdf
   \usepackage[pdftex]{graphicx}
  \else
   \usepackage[dvipdfmx]{graphicx}
  \fi
\usepackage{amssymb}
\usepackage{amsfonts}
\usepackage{bm}

\begin{document}
\title{$k$-connectivity of Random Graphs and Random Geometric Graphs in Node Fault Model} 
\author{%
  \IEEEauthorblockN{Satoshi Takabe and Tadashi Wadayama}
  \IEEEauthorblockA{Nagoya Institute of Technology,\\
                    Gokiso-cho, Showa-ku, Nagoya,
                    Aichi, Japan\\
                    Email: \{s\_takabe, wadayama\}@nitech.ac.jp}
}

\maketitle

\begin{abstract}
$k$-connectivity of random graphs is a fundamental property indicating reliability of multi-hop wireless sensor networks (WSN).
WSNs comprising of sensor nodes with limited power resources are modeled by random graphs with unreliable nodes, 
which is known as the node fault model.
In this paper, we investigate $k$-connectivity of random graphs in the node fault model by evaluating the network breakdown probability,
i.e., the disconnectivity probability of random graphs after stochastic node removals.
Using the notion of a strongly typical set, we obtain universal asymptotic upper and lower bounds of the network breakdown probability.
The bounds are applicable both to random graphs and to random geometric graphs.
We then consider three representative random graph ensembles:
the Erd\"os-R\'enyi random graph as the simplest case, the random intersection graph for WSNs with random key predistribution schemes,
and the random geometric graph as a model of WSNs generated by random sensor node deployment.
The bounds unveil the existence of the phase transition of the network breakdown probability
 for those ensembles. 
\end{abstract}

\section{Introduction}
Wireless sensor networks (WSN) have attracted a great deal of attention in various fields.
The WSN usually consists of a large number of sensor nodes with short-range radio transceivers and limited power resources.
Such an autonomous sensor node has advantages in cost efficiency, which encourages wide range of applications of WSNs in harsh environments.
Sensor nodes are expected to retain required functions of the whole network without any maintenance.
For instance, \emph{$k$-connectivity} is a crucial network property for multi-hop WSNs~\cite{Zhu}.
In practice, efficient establishment of connectivity is a central issue in design of WSNs~\cite{Al}.

Connectivity of \emph{random} graphs is a main subject for theoretical analyses of WSNs 
because their sensor nodes are usually deployed randomly, i.e., they are regarded as wireless \emph{ad hoc} networks.
Nodes and edges in graphs respectively correspond to sensor nodes and communication links in WSNs.
As an abstract model, the wireless ad hoc network is naively described by conventional random graphs~\cite{RG1} deeply investigated in random graph theory~\cite{Bol}.
Recently, however, to capture more practical aspects, other types of random graphs have been extensively studied:
\emph{random intersection graphs} as a model of secure WSNs with random key predistribution schemes~\cite{key} and
\emph{random geometric graphs} as a model of random sensor node deployment.
Analyses on those models provide theoretical evaluations of design and topological control of WSNs~\cite{Sant} to reduce
energy consumption of sensor nodes.

In addition, it is worth investigating \emph{imperfect} networks because autonomous sensor nodes and their communication links in WSNs are often at fault.
Connectivity of graphs with stochastic edge removals is first investigated by Moore and Shannon~\cite{MoS},
which has been extensively studied as connectivity of wireless ad hoc networks with accidental faults in communication links~\cite{Ball}.
The model on random intersection graphs is also studied as the robustness of secure WSNs~\cite{Zhao13}.

As another imperfect network model, the \emph{node fault model} in wireless ad hoc networks was proposed relatively recently~\cite{The}.
In the model, each node is stochastically at fault and the network breakdown probability that
the resultant graph is not $k$-connected is evaluated.
This model describes WSNs with \emph{unreliable} sensor nodes with limited power resources
or with a random wake/sleep management.

Recently, the node fault model has been studied combined with random graphs.
Nozaki \textit{et al.} studied the network breakdown probability of regular random graphs by combinatorics~\cite{nnw}.
They obtained a tight upper bound of the probability when $k=1$ while a lower bound is unrevealed.
Motivated by this work, the authors obtained an approximation formula of the probability for a wide class of random graphs~\cite{tnw}.
The formula is derived by a mean-field approximation and asymptotically agrees with numerical results.
Moreover, the asymptotic analysis of the formula predicts the existence of the phase transition
while the mathematical rigor of the results is left as an open problem.
For random geometric graphs, the model is studied using probabilistic theory,
 which proved the existence of the phase transition when $k=1$~\cite{Peng}.
Analyses in these studies are specific to either random graphs or random geometric graphs.
Providing a clear view to the node fault model is crucial to understand the phase transition in the node fault model 
and the reliability of various kinds of WSNs.

In this paper, we provide a probabilistic analysis of the node breakdown probability with respect to $k$-connectivity,
which is applicable both to random graphs and to random geometric graphs. 
Our analysis is based on asymptotic upper and lower bounds of the network breakdown probability obtained by the notion of a strongly typical set.
Those bounds enable us to universally obtain rigorous results for homogeneous random graphs defined in Sec.~\ref{main}
 including random intersection graphs and random geometric graphs.
We emphasize that our approach simplifies the proof given in~\cite{Peng} and provides general results
 with any $k$ for random geometric graphs.

This paper is organized as follows. 
In Sec.~\ref{rel}, we introduce random graphs and related results on the phase transition of $k$-connectivity.
In Sec.~\ref{def}, the node fault model and the network breakdown probability are introduced.
We describe main results on the network breakdown probability with respect to $k$-connectivity in Sec.~\ref{main}:
the asymptotic bounds are given in the first subsection,
and some rigorous results of the phase transition are provided in the remainder.
The last section is devoted to the summary of this paper.

\section{Related Works}\label{rel}
In this section, we define random graphs and $k$-connectivity, and briefly introduce some related works.

\subsection{Random Graphs}
Conventional random graphs~\cite{Bol} are defined without metric.
Let $\mathcal{G}^n$ be a set of undirected labeled simple graphs with $n$ nodes.
The random graph ensemble $\Omega_n$ is defined as a probability space $(\mathcal{G}_n,2^{\mathcal{G}_n},P)$, where
$2^{\mathcal{G}_n}$ is the power set of $\mathcal{G}_n$ and $P$ is a measure on $2^{\mathcal{G}_n}$.
The Erd\"os-R\'enyi (ER) random graph $\mathcal{G}(n,p)$ is defined
as $P(G)=(1-p)^{M-|E|}p^{|E|}$ where $M=\binom{n}{2}$, $|E|$ denotes the number of edges in $G$,
and $p\in[0,1]$ represents the probability that each pair of nodes has an edge.
A property $Q$ of graphs of order $n$ is defined as a subset of $\mathcal{G}^n$~\cite{Bol}.
For random graphs $\Omega_n$, we define $P_{\Omega_n}(Q)$ as a probability that a graph generated from $\Omega_n$ belongs to $Q$.
An ensemble $\Omega_n$ is said to have a property $Q$ asymptotically almost surely (a.a.s.)
 if $P_{\Omega_n}(Q)$ converges to one as $n\rightarrow \infty$.

Connectivity is a graph property defined as a subset of $\mathcal{G}^n$
 in which there exists a path between any pair of nodes.
For some random graphs, connectivity exhibits a \emph{phase transition} (or threshold) phenomenon~\cite{ER59, Wor}.
The threshold of ER random graphs is given as $\ln n/n$~\cite{ER59} indicating that ER random graphs have connectivity a.a.s.
 iff $p=(\ln n+c)/n$ as $c\rightarrow \infty$.
The result is naturally extended to $k$-(vertex) connectivity, a graph property that 
a graph is connected after removing any set of fewer than $k$ nodes. 
For ER random graphs, the threshold of $k$-connectivity is given as $(\ln n+(k-1)\ln\ln n)/n$~\cite{Bol},
 where that of connectivity is recovered by substituting $k=1$.

\subsection{Random Intersection Graphs}
The generalized random intersection graph is a random graph ensemble originally proposed by Singer-Cohen~\cite{rig}. 
This ensemble can be used as an abstract model of WSNs with random key predistribution schemes~\cite{grig}.
Let $\mathcal{P}=\{1,\dots,P_n\}$ be a pool of keys and $\mathcal{D}:\mathcal{P}\rightarrow[0,1]$ be a probability distribution.
A graph is then generated by the following steps: (i) for each vertex $i$, let $s_i$ be an i.i.d. random variable 
whose probability distribution is given by $\mathcal{D}$.
(ii) A subset $S_i$ of $\mathcal{P}$ is randomly generated by uniformly selecting $s_i$ different keys from $\mathcal{P}$.
(iii) An edge is assigned between two distinct nodes $i$ and $j$ iff $S_i\cap S_j\neq \emptyset$.
The resultant ensemble is called generalized random intersection graphs~\cite{grig} with $n$ nodes
 and is denoted by $\mathcal{G}(n,P_n,\mathcal{D})$.
$k$-connectivity of $\mathcal{G}(n,P_n,\mathcal{D})$ was investigated by Zhao \textit{et. al}~\cite{Zhao14} while
earlier works studied some special cases~\cite{Black,Yagan,Zhao142}.

\subsection{Random Geometric Graphs}
Random geometric graphs are defined in a metric space.
We consider a domain $S$ in the two-dimensional Euclidean space $\mathbb{R}^2$ herein.
Uniform $n$-point process $\mathcal{X}_n$ over $S$ is defined as independently uniform deployment of $n$ points in $S$.
The simplest random geometric graph $\mathcal{G}_r(\mathcal{X}_n)$~\cite{Gil} is then defined as follows: 
for a given set of points $V$ from $\mathcal{X}_n$ over $S$, edges are assigned between pairs of points whose distance is at most $r(>0)$.
Phase transitions of connectivity and $k$-connectivity of $\mathcal{G}_r(\mathcal{X}_n)$ are respectively proved in~\cite{Det1} and in~\cite{Wan}.
The phase transition threshold is known as the hitting transmission radius in the literature.
As another class of random geometric graphs, 
for instance, connectivity of an ensemble defined by the Poisson point process and a general connection function is recently studied~\cite{Guo}.
Although we can extend the arguments in the subsequent sections to such an ensemble, 
we here concentrate on the case of the simplest random geometric graphs.

\section{Model Definition}\label{def}
We define the node fault model and the network breakdown probability in this section.

The \emph{node fault model with node breakdown probability} $\epsilon$ is a stochastic process
 that, for $G\!=\!(V,E)\!\in\! \mathcal{G}^n$, each node is independently added to subset $V_b$ of {fault nodes} with probability $\epsilon$.
A \emph{survival graph} of $G$ is defined as the subgraph $G\backslash V_b$ induced by $V\backslash V_b$.
A fault node in $V_b$ can be regarded as a broken sensor node in a WSN.
Such a sensor node cannot be used for packet relaying. We thus are interested in connectivity of the survival graph.

For a random graph ensemble $\Omega_n=(\mathcal{G}^n,2^{\mathcal{G}^n},P)$, we define 
the \emph{survival graph ensemble} $\Omega_n^{(\epsilon)}$ as the resultant ensemble of survival graphs
in the {node fault model with node breakdown probability} $\epsilon$.
Note that the ensemble is defined over $\mathcal{G}_{\mathrm{s}}^n\!\triangleq\! \cup_{s=0}^n \mathcal{G}^s$.
The \emph{network breakdown probability} ${P}_{\Omega_n}^{(k)}(\epsilon)$ \emph{with respect to $k$-connectivity}
 of $\Omega_n$ with node breakdown probability $\epsilon$ is defined as the average probability
 that a survival graph generated from $\Omega_n^{(\epsilon)}$ is not $k$-connected.
Our goal is to investigate the asymptotic behavior of ${P}_{\Omega_n}^{(k)}(\epsilon)$. 

Mathematically, the survival graph ensemble is defined as a probability space.
Let $I[A]$ be an indicator function which takes one if a condition $A$ is satisfied and zero otherwise.
For $G=(V(G),E(G)),\, H=(V(H),E(H))\in \mathcal{G}^n$, $G\simeq H$ means that there exists a graph isomorphism $f:V(G)\rightarrow V(H)$
 such that $(f(u),f(v))\in E(H)$ iff $(u,v)\in E(G)$.
The survival graph ensemble $\Omega_n^{(\epsilon)}$ is then defined as $(\mathcal{G}^n_{\mathrm{s}},2^{\mathcal{G}^n_{\mathrm{s}}},Q)$
where the probability distribution $Q(\cdot)$ is given by
\begin{align}
Q(G)\triangleq &\sum_{G'\in\mathcal{G}^n} P(G') \sum_{s=0}^n (1-\kappa)^{n-s}\kappa^s \nonumber\\
\times&\sum_{V_b\in2^V; |V_b|={n-s}} I[G\simeq G'\backslash V_b],
 \label{eq_hh2}
\end{align}
for any $G$ in $\mathcal{G}^n_{\mathrm{s}}$, where $\kappa\triangleq 1-\epsilon$.

We then define a survival graph ensemble with $s$ nodes, where $s$ is an integer in $[0,n]$.
A \emph{survival graph ensemble} $\Omega_{n,s}^{(\epsilon)}$ \emph{of order} $s$ is a random graph ensemble
$(\mathcal{G}^s,2^{\mathcal{G}^s},Q(\cdot|s))$ where the probability distribution $Q(\cdot|s)$ is given by
\begin{equation}
Q(G|s) \!\triangleq\! \binom{n}{s}^{-1}\!\!\! \sum_{G'\in\mathcal{G}^n}\! P(G') 
\sum_{\substack{V_b\in2^V;\\ |V_b|={n-s}}} \!\! I[G\simeq G'\backslash V_b],
 \label{eq_hh2a}
\end{equation}
for any $G$ in $\mathcal{G}^{s}$.
We find that the distribution $Q(\cdot|s)$ is a conditional distribution of $Q(\cdot)$, i.e.,
\begin{equation}
Q(G)=\sum_{s=0}^{n}B(s;n,\kappa)Q(G|s), \label{eq_pra1}
\end{equation}
where $B(s;n,\kappa)\triangleq \binom{n}{s}(1-\kappa)^{n-s}\kappa^s$ is the binomial distribution with mean $n\kappa$.

Let us characterize the network breakdown probability ${P}_{\Omega_n}^{(k)}(\epsilon)$ averaged over $\Omega_{n}^{(\epsilon)}$
by introducing that of $\Omega_{n,s}^{(\epsilon)}$.
A \emph{network breakdown probability $P_{\Omega_n}^{(k)}(\epsilon;s)$ of $\Omega_{n,s}^{(\epsilon)}$} is defined as the probability 
that a graph generated from $\Omega_{n,s}^{(\epsilon)}$ is not $k$-connected.
By~(\ref{eq_pra1}), the breakdown probability ${P}_{\Omega_n}^{(k)}(\epsilon)$ of the whole survival graph ensemble $\Omega_{n}^{(\epsilon)}$ is given by
\begin{equation}
{P}_{\Omega_n}^{(k)}(\epsilon)=\sum_{s=0}^{n}B(s;n,\kappa)P_{\Omega_n}^{(k)}(\epsilon;s). \label{eq_pra2}
\end{equation}
Equation (\ref{eq_pra2}) is crucial to obtain asymptotic upper and lower bounds presented in the next section.


\section{Main Results}\label{main}
In this section, we describe our main results related to the phase transition of $k$-connectivity of
random graphs and random geometric graphs.

\subsection{Asymptotic Bounds}
We first present asymptotic upper and lower bounds for the network breakdown probability.
\begin{lem}\label{lem_typ_bre}
Let $\delta_n$ be a sequence for all $n$ which satisfies $\delta_n\rightarrow 0$ and $\delta_n\sqrt{n}\rightarrow \infty$
 as $n\rightarrow\infty$~\footnote{Conditions on a sequence $\delta_n$ are satisfied if, e.g., $\delta_n=n^{-1/3}$.}.
Then, for sufficiently large $n$, the network breakdown probability satisfies 
\begin{equation}
\begin{aligned}
\left(1-\frac{1}{2n\delta_n}\right)&\min_{s\in[s^{(-)}, s^{(+)}]}P_{\Omega_n}^{(k)}(\epsilon;s)
\le P_{\Omega_n}^{(k)}(\epsilon)\\
&\quad\quad\le\frac{1}{2n\delta_n}+\max_{s\in[s^{(-)}, s^{(+)}]}P_{\Omega_n}^{(k)}(\epsilon;s), 
\end{aligned}
\label{eq_typ4}
\end{equation}
where $s^{(\pm)}\triangleq (\kappa\pm\delta_n)n$.
\end{lem}

The proof is based on typicality of a strongly typical set~\cite{CK} and is available in Appendix.
It should be noted that the asymptotic bounds are applicable to an arbitrary random graphs and 
random geometric graphs defined by the uniform $n$-point process.

Combining the bounds and an exact result on the phase transition of $k$-connectivity enables us to
concisely study the asymptotic behavior of the network breakdown probability.
We specifically examine \emph{homogeneous} random graphs whose edges are i.i.d. assigned.
The ensembles studied herein are homogeneous
 while, e.g., random regular graphs and random $k$-nearest neighbor graphs are inhomogeneous
because of their degree constraints.
For a homogeneous random graph ensemble $\Omega_n$, its survival graph ensembles $\Omega_{n,s}^{(\epsilon)}$ of order $s$
are equivalent to the ensemble $\Omega_s$ with $s$ nodes.
This fact simplifies evaluations of the maximum and minimum of the network breakdown probability in~(\ref{eq_typ4}).

For inhomogeneous random graph ensembles, in contrast, the difficulty lies in the fact
 that survival graph ensembles $\Omega_{n,s}^{(\epsilon)}$ is defined as
 a nontrivial distribution different from that of the original ensemble in general.
Further analysis of $\Omega_{n,s}^{(\epsilon)}$ is necessary for the evaluation,
which is left as a future work.

\subsection{Erd\"os-R\'enyi Random Graphs}
One crucial example of homogeneous random graph ensembles is ER random graphs $\mathcal{G}(n,p)$.
The asymptotic behavior of the network breakdown probability of ER random graphs is summarized as follows.
\begin{theorem}\label{prop_M}
If $\Omega_n = \mathcal{G}(n,p)$, 
for a sequence $\omega_n$ satisfying $\omega_n\rightarrow\infty$ as $n\rightarrow\infty$ and any fixed natural number $k$, we have the following statements.
\begin{itemize}
\item[1.] If $p=(\ln n+(k-1)\ln\ln n+\omega_n)/(\kappa n)$,
then $\lim_{n\rightarrow\infty}{P}_{\Omega_n}^{(k)}(\epsilon)= 0$.
\item[2.] If $p=p^\ast\triangleq (\ln n+(k-1)\ln\ln n)/(\kappa n)$,
then $\lim_{n\rightarrow\infty}{P}_{\Omega_n}^{(k)}(\epsilon)= 1-e^{-\kappa^{1/(k-1)!}}$. 
\item[3.] If $p=(\ln n+(k-1)\ln\ln n-\omega_n)/(\kappa n)$, 
then $\lim_{n\rightarrow\infty}{P}_{\Omega_n}^{(k)}(\epsilon)= 1$. 
\end{itemize}
\end{theorem}

In the proof, we use the asymptotic property of the minimum degrees to calculate ${P}_{\Omega_n}^{(k)}(\epsilon)$.
\begin{theorem}[\cite{Bol}, Theorem 3.5]\label{thm35}
Let $m$ be any non-negative integer and $c$ be any fixed real number.
If we set $p=(\ln n+m\ln\ln n+c+o(1))/n$, we have
\begin{equation}
\begin{aligned}
&\lim_{n\rightarrow\infty}\mathrm{Pr}[\delta(G_{n,p})=m]= 1-e^{-e^{-c/m!}},\\
&\lim_{n\rightarrow\infty}\mathrm{Pr}[\delta(G_{n,p})=m+1]= e^{-e^{-c/m!}},
\end{aligned}\label{eq_pram2}
\end{equation}
where $\delta(G)$ represents the minimum degree of a graph $G$ and $G_{n,p}$ is an instance of ER random graphs $\mathcal{G}(n,p)$.
\end{theorem}


\begin{IEEEproof}[Proof of Thm.~\ref{prop_M}]
We prove the second case. The other cases can be easily proved.
Because $\Omega_{n,s}^{(\epsilon)}\!=\!\mathcal{G}(s,p)$ holds and
$G_{s,p}$ with $\delta(G_{s,p})\!=\!k$ is $k$-connected a.a.s.\cite{BT}, 
we have
${P}_{\Omega_n}^{(k)}(\epsilon;s)\!=\!\mathrm{Pr}[\delta(G_{s,p})\!<\!k]$ for any $s\in[s^{(-)}\!,s^{(+)}]$ as $n\!\rightarrow\!\infty$.

Let $\delta_n$ be a sequence that satisfies $\delta_n\ln n\rightarrow 0$ and $\delta_n\sqrt{n}\rightarrow \infty$
as $n\rightarrow\infty$.
From the definition of $s^{(\pm)}$ in Lem.~\ref{lem_typ_bre}, we find
\begin{align}
p^\ast&=\left(1\pm \frac{\delta_n}{\kappa}\right)\frac{\ln s^{(\pm)}+(k-1)\ln\ln s^{(\pm)}-\ln\kappa+o(1)}{s^{(\pm)}}\nonumber\\
&=\frac{\ln s^{(\pm)}+(k-1)\ln\ln s^{(\pm)}-\ln\kappa+o(1)}{s^{(\pm)}}.  \label{eq_pram7}
\end{align}
When $p=p^\ast$, using Thm.~\ref{thm35} and monotonicity of $k$-connectivity~\cite{Bol},
 we have $\lim_{n\rightarrow \infty}{P}_{\Omega_n}^{(k)}(\epsilon;s)= 1-e^{-\kappa^{1/(k-1)!}}$ for any $s\in[s^{(-)},s^{(+)}]$.
By Lem.~\ref{lem_typ_bre} we obtain $\lim_{n\rightarrow \infty}{P}_{\Omega_n}^{(k)}(\epsilon)= 1-e^{-\kappa^{1/(k-1)!}}$.
\end{IEEEproof}

It should be emphasized that the result when $k=1$ exactly coincides with that obtained by a mean-field approximation
 known as the cavity method in statistical physics~\cite{tnw}.
In the analysis, a random graph is approximated to a random tree with the same degree distribution.
It provides an approximation formula of the network breakdown probability for finite $n$.
For instance, for ER random graphs $\mathcal{G}(n,p)$, it reads 
\begin{equation}
P_{\Omega_n}^{\mathrm{MF}}(\epsilon)=1-\left[1-(1-\epsilon)\left\{1-p(1-\epsilon)(1-\tilde{I}_n)\right\}^{n-1}\right]^n, \label{eq_er2}
\end{equation}
where $\tilde{I}_n\in(0,1]$ is the smallest solution of 
\begin{equation}
I_n=\left[1-p(1-\epsilon)(1-I_n)\right]^{n-2}. \label{eq_er1}
\end{equation}
Its asymptotic analysis predicts a phase transition equivalent to Thm.~\ref{prop_M}.
However, the prediction is not directly proved because the rigor of the approximation is generally unrevealed.
The proof in this paper provides a strong evidence of the rigor of results obtained by the mean-field approximation.

\subsection{Random Intersection Graphs}
Another example of homogeneous random graphs is the generalized random intersection graph $\mathcal{G}(n,P_n,\mathcal{D})$ for secure WSNs.
It is apparent that the ensemble is homogeneous because of the i.i.d. key predistribution process.
Let $\mathbb{E}[X]$ and $\mathrm{Var}[X]$ respectively denote the mean and variance of a random variable $X$.

\begin{theorem}\label{prop_R}
Let $\Omega_n = \mathcal{G}(n,P_n,\mathcal{D})$ and
$X$ be a random variable following the distribution $\mathcal{D}$.
If $\mathbb{E}[X]=\Omega(\sqrt{\ln n})$ and $\mathrm{Var}[X]=o\left({\mathbb{E}[X]^2}/\{n(\ln n)^2\}\right)$ hold,
for a sequence $\omega_n$ satisfying $\omega_n=o(\ln n)$ and $\omega_n\rightarrow\infty$ as $n\rightarrow\infty$,
 we have the following statements.
\begin{itemize}
\item[1.] If ${\mathbb{E}[X]^2}/{P_n}=(\ln n+(k-1)\ln\ln n+\omega_n)/(\kappa n)$, then
$\lim_{n\rightarrow\infty}{P}_{\Omega_n}^{(k)}(\epsilon)= 0$.
\item[2.] If ${\mathbb{E}[X]^2}/{P_n}=p^\ast=(\ln n+(k-1)\ln\ln n)/(\kappa n)$, then
$\lim_{n\rightarrow\infty}{P}_{\Omega_n}^{(k)}(\epsilon)= 1-e^{-\kappa^{1/(k-1)!}}$. 
\item[3.] If ${\mathbb{E}[X]^2}/{P_n}=(\ln n+(k-1)\ln\ln n-\omega_n)/(\kappa n)$, then
$\lim_{n\rightarrow\infty}{P}_{\Omega_n}^{(k)}(\epsilon)= 1$. 
\end{itemize}
\end{theorem}
The phase transition threshold suggests that the random intersection graph is asymptotically close to the ER random graph in terms of $k$-connectivity.
It is known, however, that random intersection graphs have high clustering coefficients~\cite{Yan} implying locally dense communication links.
The proof is similar to that of Thm.~\ref{prop_M}.
We employ the following theorem instead of Thm.~\ref{thm35} in the proof.
\begin{theorem}[\cite{Zhao14}, Theorem 1]\label{thm14}
Let $X$ be a random variable following the distribution $\mathcal{D}$ and 
 $\alpha_n$ be a sequence defined by ${\mathbb{E}[X]^2}/{P_n}=(\ln n+(k-1)\ln\ln n+\alpha_n)/n$
 for all $n$.
If $\mathbb{E}[X]=\Omega(\sqrt{\ln n})$, $\mathrm{Var}[X]=o\left({\mathbb{E}[X]^2}/\{n(\ln n)^2\}\right)$,
and $|\alpha_n|=o(\ln n)$ hold,
then
\begin{align}
\lim_{n\rightarrow \infty}&\mathrm{Pr}[G(n,P_n,\mathcal{D})\mbox{ is $k$-connected}]\nonumber\\
=&
\begin{cases}
0\quad &(\lim_{n\rightarrow \infty}\alpha_n= -\infty),\\
e^{-\frac{e^{-\alpha^\ast}}{(k-1)!}}\quad &(\lim_{n\rightarrow \infty}\alpha_n= \alpha^\ast\in(-\infty,\infty)),\\
1\quad &(\lim_{n\rightarrow \infty}\alpha_n= \infty),
\end{cases}
\label{eq_z1}
\end{align}
where $G(n,P_n,\mathcal{D})$ is an instance of $\mathcal{G}(n,P_n,\mathcal{D})$.
\end{theorem}

\begin{IEEEproof}[Proof of Thm.~\ref{prop_R}]
We prove the second case. The other cases are easily proved.
Because $\Omega_{n,s}^{(\epsilon)}=\mathcal{G}(s,P_n,\mathcal{D})$ holds,
we have
${P}_{\Omega_n}^{(k)}(\epsilon;s)=1-\mathrm{Pr}[G(s,P_n,\mathcal{D})\mbox{ is $k$-connected}]$.

Letting $\delta_n$ be a sequence that satisfies $\delta_n\ln n\rightarrow 0$ and $\delta_n\sqrt{n}\rightarrow \infty$
as $n\rightarrow\infty$, (\ref{eq_pram7}) holds.
The definition of $s^{(\pm)}$ in Lem.~\ref{lem_typ_bre} indicates that $s=\Theta(n)$, which 
implies $\mathbb{E}[X]=\Omega(\sqrt{\ln s})$ and $\mathrm{Var}[X]=o\left({\mathbb{E}[X]^2}/\{s(\ln s)^2\}\right)$
for $s\in [s^{(-)},s^{(+)}]$ if $\mathbb{E}[X]=\Omega(\sqrt{\ln n})$ and $\mathrm{Var}[X]=o\left({\mathbb{E}[X]^2}/\{n(\ln n)^2\}\right)$ hold.

When ${\mathbb{E}[X]^2}/{P_n}=p^\ast$, using Thm.~\ref{thm14} and monotonicity of $k$-connectivity,
 we have $\lim_{n\rightarrow \infty}{P}_{\Omega_n}^{(k)}(\epsilon;s)= 1-e^{-\kappa^{1/(k-1)!}}$ for any $s\in[s^{(-)},s^{(+)}]$.
By Lem.~\ref{lem_typ_bre}, we have $\lim_{n\rightarrow \infty}{P}_{\Omega_n}^{(k)}(\epsilon)= 1-e^{-\kappa^{1/(k-1)!}}$.
\end{IEEEproof}

\subsection{Random Geometric Graphs}
The random geometric graph $\mathcal{G}_r(\mathcal{X}_n)$ is a homogeneous random graph ensemble 
whose nodes and edges are independently assigned.
Like the above two ensembles, it implies that the threshold of the network breakdown probability is evaluated by Lem.~\ref{lem_typ_bre}.
Letting ${G}_r(\mathcal{X}_n)$ be an instance of $\mathcal{G}_r(\mathcal{X}_n)$,
the asymptotic property of the minimum degree of ${G}_r(\mathcal{X}_n)$ is given by the following theorems.
\begin{theorem}[\cite{Det1}, Theorem 1.1]\label{thm22}
Suppose that $S$ is a unit-area square.
Then, for $\mathcal{X}_n$ over $S$, we have
\begin{equation}
\lim_{n\rightarrow\infty}\mathrm{Pr}[\delta(G_{r_n}(\mathcal{X}_n))\ge 1]= e^{-e^{-c}}, \label{eq_q05}
\end{equation}
where $r_n =\sqrt{(\ln n+c)/(\pi n)}$ for a constant $c$.
\end{theorem}
\begin{theorem}[\cite{Wan}, Theorem 2]\label{thm22a}
Suppose that $S$ is a unit-area square.
Let 
\begin{equation}
r_n =\sqrt{\frac{\ln n+(2k-1)\ln\ln n+\xi}{\pi n}}, \label{eq_q03}
\end{equation}
where 
\begin{equation}
\xi =
\begin{cases}
-2\ln\left(\sqrt{e^{-c}+\frac{\pi}{4}}-\frac{\sqrt{\pi}}{2}\right)\quad &(k= 1),\\
2\ln\frac{\sqrt{\pi}}{2^{k-1}k!}+2c\quad &(k>1),
\end{cases}
\label{eq_q04}
\end{equation}
for a constant $c$.
Then, for $\mathcal{X}_n$ over $S$, we have
\begin{equation}
\lim_{n\rightarrow\infty}\mathrm{Pr}[\delta(G_{r_n}(\mathcal{X}_n))\ge k+1]=e^{-e^{-c}}. \label{eq_q05a}
\end{equation}
\end{theorem}

Unifying these results, we obtain the following theorem in the case of the node fault model.
\begin{theorem}\label{prop_rg1} 
Suppose that $S$ is a unit-area square.
For $k\in\mathbb{N}$, let
\begin{equation}
\xi =
\begin{cases}
0\quad &(k= 1,2),\\
2\ln\frac{\sqrt{\pi}}{2^{k-2}(k-1)!}\quad &(k>2),
\end{cases}
\label{eq_q4}
\end{equation}
 $[x]_+\triangleq \max\{x,0\}$, and $\omega_n$ be a sequence such that $\omega_n\rightarrow\infty$ as $n\rightarrow \infty$. 
Then, for $\mathcal{X}_n$ over $S$ and $\Omega_n=\mathcal{G}_r(\mathcal{X}_n)$, we have the following statements.
\begin{itemize}
\item[1.] If $r=\sqrt{{(\ln n+[2k-3]_+\ln\ln n+\omega_n)}/{(\kappa \pi n)}}$, then
 $\lim_{n\rightarrow \infty}{P}_{\Omega_n}^{(k)}(\epsilon)= 0$.
\item[2.] If $r=r^\ast \triangleq\sqrt{{(\ln n+[2k-3]_+\ln\ln n+\xi)}/{(\kappa \pi n)}}$, 
then
\begin{equation}
\lim_{n\rightarrow \infty}{P}_{\Omega_n}^{(k)}(\epsilon) =
\begin{cases}
1-e^{-\kappa}\quad &(k=1),\\
1-e^{-\sqrt{\kappa(\kappa+\pi)}}\quad &(k=2),\\
1-e^{-\sqrt{\kappa}}\quad &(k\ge 3).
\end{cases}
\label{eq_q4b}
\end{equation}
\item[3.] If $r=\sqrt{{(\ln n+[2k-3]_+\ln\ln n-\omega_n)}/{(\kappa \pi n)}}$, then
 $\lim_{n\rightarrow \infty}{P}_{\Omega_n}^{(k)}(\epsilon)= 1$.
\end{itemize}
\end{theorem}

\begin{IEEEproof}
We prove the second case herein.
Because $\Omega_{n,s}^{(\epsilon)}=\mathcal{G}(s,p)$ holds and
${G}_r(\mathcal{X}_n)$ with $\delta({G}_r(\mathcal{X}_n))=k$ is $k$-connected a.a.s.~\cite{Pen}, 
we have
${P}_{\Omega_n}^{(k)}(\epsilon;s)=\mathrm{Pr}[\delta({G}_r(\mathcal{X}_s))<k]$ for any $s\in[s^{(-)},s^{(+)}]$ as $n\rightarrow \infty$.

Let $\delta_n$ be a sequence that satisfies $\delta_n\ln n\rightarrow 0$ and $\delta_n\sqrt{n}\rightarrow \infty$ as $n\rightarrow\infty$.
From the definition of $s^{(\pm)}$ in Lem.~\ref{lem_typ_bre}, we find 
\begin{equation}
r^\ast
=\frac{\ln s^{(\pm)}+[2k-3]_+\ln\ln s^{(\pm)}+\xi-\ln\kappa+o(1)}{s^{(\pm)}}.  \label{eq_pram7g}
\end{equation}
When $r=r^\ast$, using Thm.~\ref{thm22}, Thm.~\ref{thm22a}, and monotonicity of $k$-connectivity,
 we have
\begin{equation}
\lim_{n\rightarrow \infty}{P}_{\Omega_n}^{(k)}(\epsilon;s) =
\begin{cases}
1-e^{-\kappa}\quad &(k=1),\\
1-e^{-\sqrt{\kappa(\kappa+\pi)}}\quad &(k=2),\\
1-e^{-\sqrt{\kappa}}\quad &(k\ge 3),
\end{cases}
\label{eq_q4c}
\end{equation}
 for any $s\in[s^{(-)},s^{(+)}]$.
By Lem.~\ref{lem_typ_bre} we obtain (\ref{eq_q4b}).
\end{IEEEproof}

In~\cite{Peng}, the same result is given in the $k=1$ case by directly analyzing ${P}_{\Omega_n}^{(1)}(\epsilon)$
 of the survival graph ensemble $\Omega_n^{(\epsilon)}$.
It should be emphasized, however, that our analysis is based on the asymptotic bounds in Lem.~\ref{lem_typ_bre}, 
 which results in a much simpler proof when $k=1$ and general results for any $k$.

\section{Conclusion}
In this paper, we studied the network breakdown probability of random graphs and random geometric graphs
 with respect to $k$-connectivity in the node fault model.
We obtained the asymptotic upper and lower bounds for the probability, which is applicable to arbitrary random graphs.
These bounds are especially useful when a random graph ensemble is homogeneous.
Applying known results on the phase transition of $k$-connectivity,
we proved the phase transition threshold of the network breakdown probability. 
Our approach provides a clear and universal proof strategy and is potentially extensible to other random graph ensembles.

\section*{Acknowledgment}
This study is partially supported by JSPS KAKENHI Grants Numbers 16K14267, 16H02878 (TW), and 17H06758 (ST).

\appendix[Proof of Lemma~\ref{lem_typ_bre}]
Let ${T}^n_{\delta_n}$ be a strongly typical set for Bernoulli distribution $Ber(s;\kappa)\triangleq (1-\kappa)^{1-s}\kappa^s$
on $\mathcal{X}=\{0,1\}$, i.e.,
\begin{equation}
{T}^n_{\delta_n}= \{\bm{s}\in\mathcal{X}^n;|P_{\bm{s}}(a)-P(a)|\le \delta_n\,(\forall a\in\mathcal{X})\},
 \label{eq_typ5}
\end{equation}
where $P_{\bm{s}}(a)$ represents a fraction of $a$ in $\bm{s}$ and $P(a)=Ber(a;\kappa)$.
Let $s$ be the number of $1$'s in the sequence $\bm{s}$, i.e, $s\triangleq \sum_{i=1}^ns_i$.
As $P_{\bm{s}}(1)=s/n$ holds, $s\in [s^{(-)}, s^{(+)}]$ holds iff $\bm{s}\in T_{\delta_n}^n$.
Typicality of the strongly typical set~\cite[Lemma 2.12]{CK} is then represented as
\begin{equation}
1-\frac{1}{2n\delta_n}\le \mathrm{Pr.}[\bm{s}\in T_{\delta_n}^n]
=\sum_{s=s^{(-)}}^{s^{(+)}}B(s;n,\kappa)\le 1.\label{eq_typ6}
\end{equation}

From (\ref{eq_pra2}), we obtain a lower bound of the breakdown probability as follows:
\begin{align}
{P}_{\Omega_n}^{(k)}(\epsilon)
&\ge\sum_{s=s^{(-)}}^{s^{(+)}}B(s;n,\kappa)P_{\Omega_n}^{(k)}(\epsilon;s)\nonumber\\
&\ge\min_{s\in[s^{(-)}, s^{(+)}]}\{P_{\Omega_n}^{(k)}(\epsilon;s)\} \sum_{s=s^{(-)}}^{s^{(+)}}B(s;n,\kappa)\nonumber\\
&\ge\left(1-\frac{1}{2n\delta_n}\right)\min_{s\in[s^{(-)}, s^{(+)}]}P_{\Omega_n}^{(k)}(\epsilon;s).
 \label{eq_ho2}
\end{align}
In contrast, for an upper bound, we find
\begin{align}
{P}_{\Omega_n}^{(k)}(\epsilon)
&\le\frac{1}{2n\delta_n}+\sum_{s=s^{(-)}}^{s^{(+)}}B(s;n,\kappa)P_{\Omega_n}^{(k)}(\epsilon;s)\nonumber\\
&\le\frac{1}{2n\delta_n}+\max_{s\in[s^{(-)}, s^{(+)}]}\{P_{\Omega_n}^{(k)}(\epsilon;s)\} 
\sum_{s=s^{(-)}}^{s^{(+)}}B(s;n,\kappa)\nonumber\\
&\le\frac{1}{2n\delta_n}+\max_{s\in[s^{(-)}, s^{(+)}]}P_{\Omega_n}^{(k)}(\epsilon;s).
 \label{eq_ho1}
\end{align}
Combining those inequalities, we have (\ref{eq_typ4}).\hspace{\fill}\IEEEQEDclosed






\end{document}